\documentstyle[aps]{revtex}

\begin{document}
\title{remote information concentration by GHZ state and by bound entangled state}
\author{Yafei Yu$^{\thanks{%
corresponding author.}a}$ , Xiaoqing Zhou$^{a}$,Jian Feng$^{a,b}$ ,
Mingsheng Zhan$^{a}$}
\address{$^{a}$State Key Laboratory of Magnetic Resonance and Atomic and Molecular\\
Physics,\\
Wuhan Institute of Physics and Mathematics, Chinese Academy of Sciences,\\
Wuhan 430071, PR China\\
$^{b}$Institute of Optical Communication, Liaocheng University, Liaocheng ,\\
252059, Shandong, PR China}
\maketitle

\begin{abstract}
We compare remote information concentration by a maximally entangled GHZ
state with by an unlockable bound entangled state. We find that the bound
entangled state is as useful as the GHZ state, even do better than the GHZ
state in the context of communication security.

PACS number(s): 03.67.Hk
\end{abstract}

\section{Introduction}

As a peculiar feature of quantum formalism without classical counterpart,
quantum entanglement has already been at the heart of the emerging quantum
information theory. The two-qubit maximally entangled state has produced
highly nonintuitive effects such as quantum teleportation\cite{82/5/4},
quantum dense coding\cite{82/5/3}, quantum cryptography\cite{82/5/2}. In
practice, one usually deals with noisy entanglement represented by mixed
state of a composite system. All the noisy entanglement of two-qubit system
can be distilled to the singlet form\cite{82/5/11}. But, for the mixed
entangled states of more-than-two-qubit systems, there are two qualitatively
different kinds of entanglement : free entanglement which can always be
distilled and bound one which cannot be brought to the singlet form only by
local quantum operations and classical communication (LOCC)\cite{80/24}. In
the last few years, the bound entangled states of the multipartite systems
is under extensive research because it arouses a deeper understanding of the
entanglement and the nonlocality of quantum states\cite{88,65,87,66}.

On the other hand, the bound entanglement seems to be useless for quantum
information work such as reliable transmission of quantum data via
teleportation\cite{80/24,59/1}. However, it is not the case. Resent research
has shown that in a sense many copies of a bound entangled state can be
pumped into a free entangled state\cite{82/5}. And it has been found that
entanglement can be distilled from two copies of a bound entangled state
distributed to different parties\cite{shor}. Namely, even a small amount of
bound entanglement can be activated\cite{62} and used to process quantum
information\cite{88/24}. More recently, it has been demonstrated that a
single copy of a bound entangled state can perform information work better
than classically correlated states\cite{86/2}. That is, a single bound
entangled state alone is useful for quantum information processing. Here we
give an example to demonstrate that in a sense a single bound entangled
state performs as good as the maximally entangled GHZ state.

In this paper we consider that quantum information originally from a single
qubit, but now asymmetrically distributed into three spatially separated
qubits, is remote concentrated back to a single qubit by an initially shared
four-partite GHZ state without performing any global operations. This
quantum information work can also be achieved by a initially shared
four-partite unlockable bound entangled state\cite{63/032306}. We discuss
remote information concentration in detail by GHZ state in Sec.II and by the
bound entangled state in Sec.III, then give our discussion and conclusion in
Sec.IV.

\section{Remote information concentration by GHZ state}

For simplicity, we focus on the protocol of three parties.

Supposing that three separate parties Alice, Bob and Charlie hold three
qubits A, B and C, respectively. The three qubits are in an asymmetric
telecloning state 
\begin{equation}
\left| \psi \right\rangle _{ABC}=\alpha \left| \phi _{0}\right\rangle
_{ABC}+\beta \left| \phi _{1}\right\rangle _{ABC},
\end{equation}
where $\alpha $ and $\beta $ are real numbers and satisfies $\alpha
^{2}+\beta ^{2}=1$. The states $\left| \phi _{0}\right\rangle _{ABC}$ and $%
\left| \phi _{1}\right\rangle _{ABC}$ is defined as 
\begin{eqnarray}
\left| \phi _{0}\right\rangle _{ABC} &=&\frac{1}{\sqrt{N}}(\left|
0\right\rangle _{A}\left| 0\right\rangle _{B}\left| 0\right\rangle
_{C}+p\left| 1\right\rangle _{A}\left| 0\right\rangle _{B}\left|
1\right\rangle _{C}+q\left| 1\right\rangle _{A}\left| 1\right\rangle
_{B}\left| 0\right\rangle _{C}), \\
\left| \phi _{1}\right\rangle _{ABC} &=&\frac{1}{\sqrt{N}}(\left|
1\right\rangle _{A}\left| 1\right\rangle _{B}\left| 1\right\rangle
_{C}+p\left| 0\right\rangle _{A}\left| 1\right\rangle _{B}\left|
0\right\rangle _{C}+q\left| 0\right\rangle _{A}\left| 0\right\rangle
_{B}\left| 1\right\rangle _{C}),  \nonumber
\end{eqnarray}
where $q=1-p,p>q,N$ is a normalization factor given by $N=1+p^{2}+q^{2}$.
That is, the information of an unknown state $\left| \chi \right\rangle
=\alpha \left| 0\right\rangle +\beta \left| 1\right\rangle $ is diluted to a
composite system consisting of qubits A, B and C by an asymmetric
telecloning process\cite{59/156,13}. Now the task is to concentrate the
information back to a single qubit without the collective operations between
the three qubits. The schematic picture of the remote information
concentration is illustrated in Fig.1.

The first protocol we proposed is to share in advance a four-partite GHZ
state 
\begin{equation}
\left| \varphi \right\rangle _{DEFG}=\frac{1}{\sqrt{2}}(\left|
0\right\rangle _{D}\left| 0\right\rangle _{E}\left| 0\right\rangle
_{F}\left| 0\right\rangle _{G}+\left| 1\right\rangle _{D}\left|
1\right\rangle _{E}\left| 1\right\rangle _{F}\left| 1\right\rangle _{G})
\end{equation}
between three parties Alice, Bob and Charlie, which hold the qubits E, F and
G, respectively. Let David hold the qubit D. Then the three parties Alice,
Bob and Charlie perform the Bell-state measurements (BSMs) on the respective
pairs of qubits, and inform David of the results of the measurements. Each
of the three parties gets one of four possible outputs $\left\{ \left| \Phi
^{i}\right\rangle _{(i=0,1,2,3)}\right\} $ of the BSM, where $\left| \Phi
^{i}\right\rangle _{(i=0,1,2,3)}$ represent the four Bell states $\left|
\Phi ^{0,1}\right\rangle =\left( \left| 00\right\rangle \pm \left|
11\right\rangle \right) /\sqrt{2}$, $\left| \Phi ^{2,3}\right\rangle =\left(
\left| 01\right\rangle \pm \left| 01\right\rangle \right) /\sqrt{2}$. Each
of the four possible outputs is associated with the corresponding Pauli
operators $\left\{ \sigma _{(i=0,1,2,3)}^{i}\right\} $, where $\sigma
_{0}=I,\sigma _{1}=\sigma _{z},\sigma _{2}=\sigma _{x},\sigma _{3}=\sigma
_{y}$. According to their results of the measurements David rotates his
qubit D, finally obtains the state $\left| \chi \right\rangle $ on the qubit
D.

The information concentration can be understood in a formular way. We write
the states $\left| \psi \right\rangle _{ABC}$ and $\left| \varphi
\right\rangle _{DEFG}$ together: 
\begin{eqnarray}
\left| \psi \right\rangle _{ABC}\left| \varphi \right\rangle _{DEFG}
&=&\left( \alpha \left| \phi _{0}\right\rangle _{ABC}+\beta \left| \phi
_{1}\right\rangle _{ABC}\right) \otimes \frac{1}{\sqrt{2}}(\left|
0\right\rangle _{D}\left| 0\right\rangle _{E}\left| 0\right\rangle
_{F}\left| 0\right\rangle _{G}+\left| 1\right\rangle _{D}\left|
1\right\rangle _{E}\left| 1\right\rangle _{F}\left| 1\right\rangle _{G}) 
\nonumber \\
&=&\frac{1}{\sqrt{2N}}\{\alpha (\left| 00\right\rangle _{AE}\left|
00\right\rangle _{BF}\left| 00\right\rangle _{CG}\left| 0\right\rangle
_{D}+p\left| 10\right\rangle _{AE}\left| 00\right\rangle _{BF}\left|
10\right\rangle _{CG}\left| 0\right\rangle _{D}+q\left| 10\right\rangle
_{AE}\left| 10\right\rangle _{BF}\left| 00\right\rangle _{CG}\left|
0\right\rangle _{D}  \nonumber \\
&&+\left| 01\right\rangle _{AE}\left| 01\right\rangle _{BF}\left|
01\right\rangle _{CG}\left| 1\right\rangle _{D}+p\left| 11\right\rangle
_{AE}\left| 01\right\rangle _{BF}\left| 11\right\rangle _{CG}\left|
1\right\rangle _{D}+q\left| 11\right\rangle _{AE}\left| 11\right\rangle
_{BF}\left| 01\right\rangle _{CG}\left| 1\right\rangle _{D})  \nonumber \\
&&+\beta (\left| 10\right\rangle _{AE}\left| 10\right\rangle _{BF}\left|
10\right\rangle _{CG}\left| 0\right\rangle _{D}+q\left| 00\right\rangle
_{AE}\left| 00\right\rangle _{BF}\left| 10\right\rangle _{CG}\left|
0\right\rangle _{D}+p\left| 00\right\rangle _{AE}\left| 10\right\rangle
_{BF}\left| 00\right\rangle _{CG}\left| 0\right\rangle _{D}  \nonumber \\
&&+\left| 11\right\rangle _{AE}\left| 11\right\rangle _{BF}\left|
11\right\rangle _{CG}\left| 1\right\rangle _{D}+q\left| 01\right\rangle
_{AE}\left| 01\right\rangle _{BF}\left| 11\right\rangle _{CG}\left|
1\right\rangle _{D}+p\left| 01\right\rangle _{AE}\left| 11\right\rangle
_{BF}\left| 01\right\rangle _{CG}\left| 1\right\rangle _{D})\}.
\end{eqnarray}
A lengthy but straightward calculation gives that if the combining result of
the three BSMs is one of the set $\{\Phi _{AE}^{0}\Phi _{BF}^{0}\Phi
_{CG}^{0},\Phi _{AE}^{0}\Phi _{BF}^{1}\Phi _{CG}^{1},\Phi _{AE}^{1}\Phi
_{BF}^{0}\Phi _{CG}^{1},\Phi _{AE}^{1}\Phi _{BF}^{1}\Phi _{CG}^{0},\Phi
_{AE}^{2}\Phi _{BF}^{0}\Phi _{CG}^{2},\Phi _{AE}^{3}\Phi _{BF}^{0}\Phi
_{CG}^{3},\Phi _{AE}^{3}\Phi _{BF}^{1}\Phi _{CG}^{2},$

$\Phi _{AE}^{2}\Phi _{BF}^{1}\Phi _{CG}^{3},\Phi _{AE}^{2}\Phi _{BF}^{2}\Phi
_{CG}^{0},\Phi _{AE}^{3}\Phi _{BF}^{3}\Phi _{CG}^{0},\Phi _{AE}^{2}\Phi
_{BF}^{3}\Phi _{CG}^{1},\Phi _{AE}^{3}\Phi _{BF}^{2}\Phi _{CG}^{1}\}$, which
are associated with the Pauli operator $\sigma _{0}=I$, the state of the
qubit D is projected to the state $\left| \chi \right\rangle =\alpha \left|
0\right\rangle +\beta \left| 1\right\rangle $. Analogously, there is a Pauli
operator $\sigma ^{i}$ pertaining to the combining result of the three BSMs
from each of other sets. The Pauli operator $\sigma ^{i}$ is the product (up
to a global phase) of the respective Pauli operators pertaining to the three
BSMs. That is, up a global phase factor, $\sigma ^{i}$ is equal to $\sigma
_{AE}^{l}\sigma _{BF}^{j}\sigma _{CG}^{k}$, where $l,j,k=0,1,2,3$ and the
subscripts denote the BSMs on the corresponding pairs of qubits. Then David
performs the operator $\sigma ^{i}$ on his qubit D to retrieve the state $%
\left| \chi \right\rangle $. So the protocol achieves the task: $\left| \psi
\right\rangle _{ABC}\rightarrow \left| \chi \right\rangle $ with certainty
by pre-sharing a four-partite GHZ state. The task of remote information
concentration in \cite{86/2} is the special case when $p=q=\frac{1}{2}$.

It is worth noting that the distribution probability of the combining
results of the three BSMs reveals some information of the state $\left| \psi
\right\rangle _{ABC}$. For instance, while we get $\Phi _{AE}^{0}\Phi
_{BF}^{0}\Phi _{CG}^{0}$ with the probability of $\frac{1}{16N}$, we obtain $%
\Phi _{AE}^{2}\Phi _{BF}^{0}\Phi _{CG}^{2}$ with the probability of $\frac{%
p^{2}}{16N}$ and $\Phi _{AE}^{2}\Phi _{BF}^{2}\Phi _{CG}^{0}$ with the
probability of $\frac{q^{2}}{16N}$.

Consequently, by utilizing a pre-sharing four-partite GHZ state a remote
information concentration by LOCC can be achieved with certainty. And the
distribution of the combing results of the three BSMs reveals some
information of the state $\left| \psi \right\rangle _{ABC}$.

\section{Remote information concentration by bound entangled state}

In the above discussion a remote information concentration is achieved by
sharing a GHZ state and LOCC. In the following we demonstrate that the
information work can also be implemented by an unlockable bound entangled
state and LOCC.

In order to concentrate the information diluted in the state $\left| \psi
\right\rangle _{ABC}$ back to the state $\left| \chi \right\rangle $ on a
single qubit, a four-partite unlockable bound entangled state 
\begin{equation}
\rho _{DEFG}^{ub}=\frac{1}{4}\sum_{i=0}^{3}\left| \Phi _{i}\right\rangle
_{DE}\left\langle \Phi _{i}\right| \otimes \left| \Phi _{i}\right\rangle
_{FG}\left\langle \Phi _{i}\right| ,
\end{equation}
where $\left| \Phi _{i}\right\rangle $ is defined as before, is pre-shared
between Alice, Bob, Charlie and David. In the same way as above, the qubit E
is sent to Alice, the qubit F to Bob, the qubit G to Charlie and the qubit D
to David. No joint operations between qubits belonging to different parties
is allowed. The three parties Alice, Bob, and Charlie perform the Bell-state
measurements on their respective pairs of qubits in hand. Likewise, each of
them obtains one of the possible outputs $\left\{ \left| \Phi
^{i}\right\rangle _{(i=0,1,2,3)}\right\} $ of the BSM, which is associated
with a corresponding Pauli operator in the set $\left\{ \sigma
_{(i=0,1,2,3)}^{i}\right\} $, and communicates the result with David,
respectively. David determines a Pauli operator $\sigma ^{i}$ on his qubit D
for retrieving the state $\left| \chi \right\rangle $ on the qubit D,
according the product of the three Pauli operators pertaining to each one of
the three BSMs. Finally, the pure state $\left| \chi \right\rangle $ comes
out on the qubit D.

The process also can be expressed in a formular way. We joint the state $%
\left| \psi \right\rangle _{ABC}$ with $\rho _{DEFG}^{ub}$: 
\begin{eqnarray}
\left| \psi \right\rangle _{ABC}\otimes \rho _{DEFG}^{ub}\otimes
_{ABC}\left\langle \psi \right| &=&(\alpha \left| \phi _{0}\right\rangle
_{ABC}+\beta \left| \phi _{1}\right\rangle _{ABC})\rho _{DEFG}^{ub}(\alpha
_{ABC}^{*}\left\langle \phi _{0}\right| +\beta _{ABC}^{*}\left\langle \phi
_{1}\right| )  \nonumber \\
&\rightarrow &\frac{1}{64}\{(\left| I\right\rangle \left\langle I\right|
)_{AE,BF,CG}\sigma _{0}(\alpha \left| 0\right\rangle _{D}+\beta \left|
1\right\rangle _{D})(\alpha _{D}^{*}\left\langle 0\right| +\beta
_{D}^{*}\left\langle 1\right| )\sigma _{0}  \nonumber \\
&&+(\left| II\right\rangle \left\langle II\right| )_{AE,BF,CG}\sigma
_{1}(\alpha \left| 0\right\rangle _{D}+\beta \left| 1\right\rangle
_{D})(\alpha _{D}^{*}\left\langle 0\right| +\beta _{D}^{*}\left\langle
1\right| )\sigma _{1}  \nonumber \\
&&+(\left| III\right\rangle \left\langle III\right| )_{AE,BF,CG}\sigma
_{2}(\alpha \left| 0\right\rangle _{D}+\beta \left| 1\right\rangle
_{D})(\alpha _{D}^{*}\left\langle 0\right| +\beta _{D}^{*}\left\langle
1\right| )\sigma _{2}  \nonumber \\
&&+(\left| IV\right\rangle \left\langle IV\right| )_{AE,BF,CG}\sigma
_{3}(\alpha \left| 0\right\rangle _{D}+\beta \left| 1\right\rangle
_{D})(\alpha _{D}^{*}\left\langle 0\right| +\beta _{D}^{*}\left\langle
1\right| )\sigma _{3}\},
\end{eqnarray}
where$\left| I\right\rangle \left\langle I\right| $ marks a set of the
combining results of the three BSMs as 
\begin{eqnarray}
\left| I\right\rangle \left\langle I\right| &=&\left| \Phi ^{0}\right\rangle
_{AE}\left\langle \Phi ^{0}\right| \otimes \left| \Phi ^{0}\right\rangle
_{BF}\left\langle \Phi ^{0}\right| \otimes \left| \Phi ^{0}\right\rangle
_{CG}\left\langle \Phi ^{0}\right| +\left| \Phi ^{0}\right\rangle
_{AE}\left\langle \Phi ^{0}\right| \otimes \left| \Phi ^{1}\right\rangle
_{BF}\left\langle \Phi ^{1}\right| \otimes \left| \Phi ^{1}\right\rangle
_{CG}\left\langle \Phi ^{1}\right|  \nonumber \\
&&+\left| \Phi ^{1}\right\rangle _{AE}\left\langle \Phi ^{1}\right| \otimes
\left| \Phi ^{0}\right\rangle _{BF}\left\langle \Phi ^{0}\right| \otimes
\left| \Phi ^{1}\right\rangle _{CG}\left\langle \Phi ^{1}\right| +\left|
\Phi ^{1}\right\rangle _{AE}\left\langle \Phi ^{1}\right| \otimes \left|
\Phi ^{1}\right\rangle _{BF}\left\langle \Phi ^{1}\right| \otimes \left|
\Phi ^{0}\right\rangle _{CG}\left\langle \Phi ^{0}\right|  \nonumber \\
&&+\left| \Phi ^{2}\right\rangle _{AE}\left\langle \Phi ^{2}\right| \otimes
\left| \Phi ^{2}\right\rangle _{BF}\left\langle \Phi ^{2}\right| \otimes
\left| \Phi ^{0}\right\rangle _{CG}\left\langle \Phi ^{0}\right| +\left|
\Phi ^{2}\right\rangle _{AE}\left\langle \Phi ^{2}\right| \otimes \left|
\Phi ^{3}\right\rangle _{BF}\left\langle \Phi ^{3}\right| \otimes \left|
\Phi ^{1}\right\rangle _{CG}\left\langle \Phi ^{1}\right|  \nonumber \\
&&+\left| \Phi ^{3}\right\rangle _{AE}\left\langle \Phi ^{3}\right| \otimes
\left| \Phi ^{2}\right\rangle _{BF}\left\langle \Phi ^{2}\right| \otimes
\left| \Phi ^{1}\right\rangle _{CG}\left\langle \Phi ^{1}\right| +\left|
\Phi ^{3}\right\rangle _{AE}\left\langle \Phi ^{3}\right| \otimes \left|
\Phi ^{3}\right\rangle _{BF}\left\langle \Phi ^{3}\right| \otimes \left|
\Phi ^{0}\right\rangle _{CG}\left\langle \Phi ^{0}\right|  \nonumber \\
&&+\left| \Phi ^{0}\right\rangle _{AE}\left\langle \Phi ^{0}\right| \otimes
\left| \Phi ^{2}\right\rangle _{BF}\left\langle \Phi ^{2}\right| \otimes
\left| \Phi ^{2}\right\rangle _{CG}\left\langle \Phi ^{2}\right| +\left|
\Phi ^{0}\right\rangle _{AE}\left\langle \Phi ^{0}\right| \otimes \left|
\Phi ^{3}\right\rangle _{BF}\left\langle \Phi ^{3}\right| \otimes \left|
\Phi ^{3}\right\rangle _{CG}\left\langle \Phi ^{3}\right|  \nonumber \\
&&+\left| \Phi ^{1}\right\rangle _{AE}\left\langle \Phi ^{1}\right| \otimes
\left| \Phi ^{2}\right\rangle _{BF}\left\langle \Phi ^{2}\right| \otimes
\left| \Phi ^{3}\right\rangle _{CG}\left\langle \Phi ^{3}\right| +\left|
\Phi ^{1}\right\rangle _{AE}\left\langle \Phi ^{1}\right| \otimes \left|
\Phi ^{3}\right\rangle _{BF}\left\langle \Phi ^{3}\right| \otimes \left|
\Phi ^{2}\right\rangle _{CG}\left\langle \Phi ^{2}\right|  \nonumber \\
&&+\left| \Phi ^{2}\right\rangle _{AE}\left\langle \Phi ^{2}\right| \otimes
\left| \Phi ^{0}\right\rangle _{BF}\left\langle \Phi ^{0}\right| \otimes
\left| \Phi ^{2}\right\rangle _{CG}\left\langle \Phi ^{2}\right| +\left|
\Phi ^{2}\right\rangle _{AE}\left\langle \Phi ^{2}\right| \otimes \left|
\Phi ^{1}\right\rangle _{BF}\left\langle \Phi ^{1}\right| \otimes \left|
\Phi ^{3}\right\rangle _{CG}\left\langle \Phi ^{3}\right|  \nonumber \\
&&+\left| \Phi ^{3}\right\rangle _{AE}\left\langle \Phi ^{3}\right| \otimes
\left| \Phi ^{0}\right\rangle _{BF}\left\langle \Phi ^{0}\right| \otimes
\left| \Phi ^{3}\right\rangle _{CG}\left\langle \Phi ^{3}\right| +\left|
\Phi ^{3}\right\rangle _{AE}\left\langle \Phi ^{3}\right| \otimes \left|
\Phi ^{1}\right\rangle _{BF}\left\langle \Phi ^{1}\right| \otimes \left|
\Phi ^{2}\right\rangle _{CG}\left\langle \Phi ^{2}\right| .
\end{eqnarray}
It is clear that the combing results of the three BSMs $\left|
I\right\rangle \left\langle I\right| $ correspond to the case that the
product $\sigma _{AE}^{l}\sigma _{BF}^{j}\sigma _{CG(l,j,k=0,1,2,3)}^{k}$ of
the three Pauli operators pertaining to each one of the three BSMs amounts
to $\sigma ^{0}$. So the state of the qubit D is projected to the state $%
\left| \chi \right\rangle $. The rest may be deduced by analogy that $\left|
II\right\rangle \left\langle II\right| $ corresponds to the product of $%
\sigma ^{1}$, $\left| III\right\rangle \left\langle III\right| $ to the
product of $\sigma ^{2}$, $\left| IV\right\rangle \left\langle IV\right| $
to the product of $\sigma ^{3}$. Correspondingly, David does a Pauli
operator $\sigma ^{1}$, $\sigma ^{2}$ and $\sigma ^{3}$ on his qubit D in
order to get the correct state $\left| \chi \right\rangle $, respectively.
In the end, David gets a qubit D in the state $\left| \chi \right\rangle $
with certainty.

Hence, with exploiting an unlockable bound entangled state the same remote
information concentration is achieved only by LOCC as with a maximally
entangled GHZ state.

However, in the protocol with the bound entangled state, the distribution of
the combining results donot reveal any information of the state $\left| \psi
\right\rangle _{ABC}$. A complicated calculation displays that each of all
combining results of the three BSMs comes with the equal probability of $%
\frac{1}{64}$. That is, taken for example, the combining result $\Phi
_{AE}^{0}\Phi _{BF}^{0}\Phi _{CG}^{0}$ comes out with the probability of $%
\frac{1}{64}$, the other combining results $\Phi _{AE}^{2}\Phi _{BF}^{0}\Phi
_{CG}^{2}$ and $\Phi _{AE}^{2}\Phi _{BF}^{2}\Phi _{CG}^{0}$ do so. Allowing
for communication security, the unlockable bound entangled state is more
suitable for the remote information concentration than the maximally
entangled GHZ state.

\section{Conclusion}

In the paper, we consider a quantum information work, where the quantum
information initially in the state $\left| \chi \right\rangle $ of a single
qubit, but now distributed into three spatially separated qubits, is
remotely concentrated back in the state $\left| \chi \right\rangle $ of a
single qubit. It is found that a maximally entangled GHZ state and an
unlockable bound entangled state all can do the work with LOCC. It gives a
demonstration that in a sense, a single copy of a bound entangled state is
still as useful in quantum information processing as a maximally entangled
GHZ state. Even in the view of communication security, the bound entangled
state does the work better than the GHZ state, because for the former the
distribution of the combining results of the three local BSMs does not
reveal any information about the input state $\left| \psi \right\rangle
_{ABC}$.

As analyzed in \cite{86/2}, it is the entanglement existing in the state $%
\left| \psi \right\rangle _{ABC}$ that liberates the bound entangled state
for transmitting the quantum information. The entanglement in the error
correction state $\left| \psi \right\rangle _{ABC}$ is crucial for the
remote information concentration with the bound entangled state and LOCC. It
evokes another question about to what extent the bound entangled state
cannot work for this information work. The consideration is helpful for deep
understanding the relation between the free entanglement and the bound one,
and quantum entanglement itself. We hope that our work will stimulate more
research into the nature of entanglement and its applications in the quantum
communication.

\begin{center}
{\bf Acknowledgments}
\end{center}

This work has been financially supported by the National Natural Science
Foundation of China under the Grant No.10074072.

\end{document}